\documentclass[sigconf]{acmart}

\newcommand{\revision}[1]{\textcolor{black}{#1}}

\AtBeginDocument{%
  }

\copyrightyear{2025}
\acmYear{2025}
\setcopyright{cc}
\setcctype{by-nd}
\acmConference[CHI '25]{CHI Conference on Human Factors in Computing Systems}{April 26-May 1, 2025}{Yokohama, Japan}
\acmBooktitle{CHI Conference on Human Factors in Computing Systems (CHI '25), April 26-May 1, 2025, Yokohama, Japan}
\acmDOI{10.1145/3706598.3713264}
\acmISBN{979-8-4007-1394-1/25/04}

\begin{document}

\title{"Why do we do this?": Moral Stress and The Affective Experience of Ethics in Practice}

\author{Sonja Rattay}
\email{srr@di.ku.dk}
\orcid{0000-0002-8161-851X}
\affiliation{%
\institution{University of Copenhagen}
\city{Copenhagen}
\country{Denmark}\\
\institution{ITU Linz}
\city{Linz}
\country{Austria}
}

\author{Ville Vakkuri}
\orcid{0000-0002-1550-1110}
\email{ville.vakkuri@uwasa.fi}
\affiliation{%
\institution{University of Vaasa}
\city{Vaasa}
\country{Finland}
}

\author{Marco Rozendaal}
\orcid{0000-0003-4085-8602}
\email{m.c.rozendaal@tudelft.nl}
\affiliation{%
 \institution{TU Delft}
 \city{Delft}
 \country{Netherlands}
}

\author{Irina Shklovski}
\orcid{0000-0003-1874-0958}
\email{ias@di.ku.dk}
\affiliation{%
\institution{University of Copenhagen}
\city{Copenhagen}
\country{Denmark},\\
\institution{Linköping University}
\city{Linköping}
\country{Sweden}
}

\renewcommand{\shortauthors}{Rattay et al.}

\begin{abstract}
A plethora of toolkits, checklists, and workshops have been developed to bridge the well-documented gap between AI ethics principles and practice. Yet little is known about effects of such interventions on practitioners. We conducted an ethnographic investigation in a major European city organization that developed and works to integrate an ethics toolkit into city operations. We find that the integration of ethics tools by technical teams destabilises their boundaries, roles, and mandates around responsibilities and decisions. This lead to emotional discomfort and feelings of vulnerability, which neither toolkit designers nor the organization had accounted for. We leverage the concept of moral stress to argue that this affective experience is a core challenge to the successful integration of ethics tools in technical practice. Even in this best case scenario, organisational structures were not able to deal with moral stress that resulted from attempts to implement responsible technology development practices.

\end{abstract}



\received{20 February 2007}
\received[revised]{12 March 2009}
\received[accepted]{5 June 2009}

\maketitle

\section{Introduction}

Debates about ethics in computing are decidedly not new, but the proliferation of complex data-driven systems has resulted in an explosion of scholarly work and regulatory efforts. Research has circled around defining high-level ethical principles \cite{fjeld2020principled, floridi2018ai4people}, shifting between a focus on ethics and a discussion of values \cite{friedman2019value} and developing a range of guidelines \cite{correa2023worldwide, hagendorff2020ethics, jobin2019global} and codes \cite{boddington2017towards}. These efforts fostered development of principles-based checklists \cite{wong2020beyond}, toolkits \cite{morley2020initial}, workshops \cite{ballard2019judgment, stark2021apologos, taylor2020constructing}, and a variety of materials designed to support reflection in technology design and development \cite{cowls2019designing, gray2022practitioner, gray2023scaffolding}. The multitude of publications and materials mainly highlight that so far, there is little consensus on how to reliably address ethical challenges posed by rapid technological progress. 

Where such efforts have been studied, the prominent finding has been a lack of impact \cite{mcnamara2018does, winkler2021twenty} and warnings that principles, while important, are difficult to effectively translate into practice \cite{mittelstadt2019principles}. Researchers have investigated the difficulties faced by practitioners, highlighting a mismatch between expectations of operationalised ethics and the messy reality of everyday work life \cite{chivukula2020dimensions, gray2019ethical, lindberg2023cultivating, widder2023dislocated, widder2023s}. Yet despite the emerging emphasis on understanding how ethics is done in practice \cite{wong2021tactics, powell2022addressing, madaio_co-designing_2020, lindberg2023cultivating}, little research has been able to investigate how integration of ethics toolkits and other interventions may impact technological development practices long term. 

In this study we ask how do practitioners experience the implementation of an ethics toolkit in an everyday organisational context? We conducted an ethnographic study with a technical team in a major European city. This complex public services organization has already developed its own internal principles-based ethics toolkit, integrating it into city operations that focus on the design, development, and implementation of data driven technologies (we will use the term “design of AI systems” to refer to this process throughout the paper). Rather than evaluating the toolkit itself, we focus on what such an implementation does to the practitioners that take it on board. We find that practitioners in our study must find ways to cope with uncomfortable feelings, such as anxiety and frustration, that arise from identifying and engaging with challenging moral situations that become more visible as they use the toolkit. We leverage the concept of moral stress \cite{reynolds2015recognition} from moral psychology to analyze this affective experience of ethics in practice and to understand the emotional labor that affects the outcomes of ethics toolkit implementation in organisations. We identify three relational configurations relevant for the affective experience of ethics in practice – within team, beyond team, and beyond organization and showcase how moral stress unfolds differently across these.

The paper makes the following contributions: 

1. We demonstrate how the integration of ethics tools destabilises established boundaries, roles, and mandates around responsibilities and decisions in an organizational context, introducing greater uncertainty about what constitutes the right action.

2. We introduce the concept of moral stress as an unavoidable yet completely overlooked affective cost of ethics interventions for practitioners, which can explain why so many efforts towards responsible technology development fail to reach intended outcomes.

3. We illustrate the role of organizational structures and the importance of well-functioning teams as key determinants for the design of effective interventions towards responsible technology development


\section{Related Work}

\subsection{The Theory-Practice Gap in AI Ethics}
The gap between AI ethics principles and practice is well documented \cite{fjeld2020principled, mittelstadt2019principles} and there is consensus that crossing this gap is a non-trivial problem \cite{munn2023uselessness}. For example, in a review of 200 AI ethics guidelines and recommendations, Corrêa et al. \cite{correa2023worldwide} find that prescriptive normative claims are typically presented without considerations for how to achieve them. While there seems to be a convergence around which principles are most important, the interpretation and justifications of why and how these principles matter diverge widely \cite{jobin2019global}. Even the ACM code of ethics, while a decidedly important document, has little apparent impact on professionals in the tech community \cite{mcnamara2018does}. In a scoping review of responsible AI guidelines, Sadek et al. \cite{sadek2024challenges} note that the abstract nature of the principles and the lack of clear implementation procedures are important reasons for why the gap between principles and practice persists. 

In parallel to prescriptive principle-based approaches, we see a variety of materials targeting ethical concerns in technical practice, such as games \cite{ballard2019judgment, ali_ai_2023}, card-sets \cite{tkautz_cards_2021, artefact_tarot_nodate, calderon_blindspots_2021, ECCOLA}, toolkits \cite{tangible_ethical_nodate, gispen_ethics_2017, zou_design_2018}, and workshops \cite{33a_ai_2022, doteveryone_consequence_nodate, open_data_2021, hyper_island_unintended_nodate} that have been produced both by researchers and industry actors to operationalize ethical considerations into development processes. In the following we refer to such materials as ethics interventions. Reviews of these interventions consistently show that they tend to fall short of impact in the AI industry \cite{chivukula2021surveying, wong2023seeing, hagendorff2020ethics} as they do not account for the environments in which such tech practices are situated \cite{yildirim2023investigating, liao2020questioning, chivukula2020dimensions, winkler2021twenty}. For example, Morley and colleagues \cite{morley2020initial} demonstrate that almost all existing translational tools and methods are either too flexible (vulnerable to ethics washing) or too strict (unresponsive to context). The decoupling of policies, practices, and outcomes leads to practitioners facing major hurdles when trying to integrate AI ethics practices into development processes and organizational structures \cite{ali_ai_2023}. Interviews with industry leaders in particular report that lacking awareness and feasible practices are a challenge \cite{lindberg2024doing}, and that ethics efforts are perceived to be in tension with industry structures and values such as technological solutionism and market fundamentalism \cite{lindberg2024doing, metcalf2019owning}.

\subsection{Ethics as a social and relational practice}
Where principles-based approaches often take their departure in philosophical considerations of ethical concerns, studies of ethics in practice in technical contexts demonstrate that technical practitioners are already ethically engaged through other means \cite{dindler2022engagements, widder2023s, wong2021tactics, ibanez2022operationalising, taylor2020constructing, deng_investigating_2023}. Shilton’s \cite{shilton2013values} notion of value levers for example describes how organizational and operational structures in design teams and labs influence the way values and ethical deliberations are being thematized, normalized, and included or excluded in daily routines. Shilton described different infrastructural aspects that support and embed discussions about values in the design of technology, calling attention to the fact that in a market driven design field, the constant pressure of technical innovation makes it more difficult for teams to make the time for a slow and deliberate value-driven design process. More recently, Lindberg, Karlström and Barbutiu \cite{lindberg2023cultivating} report on the persistent organisational barriers that make it difficult to create emotional buy in from practitioners to invest time and resources on ethics efforts, in particular where such efforts take more complex forms to encourage practitioners to reflect and engage with ethical implications of their work beyond "tick box exercises". 

Such research highlights that ethics in practice is not the same as ethics as normative inquiry. The modularized nature of the software development workflow, inherent to any AI development, makes it difficult to pin down where exactly accountabilities and responsibilities for principles and values are located among the developers and other stakeholders \cite{widder2023dislocated}. Organisational processes such as project planning, resource management, and team hierarchies strongly influence ethical decision-making dynamics \cite{devon_design_2004}. The central debate in tech ethics is therefore not whether or which ethics is desirable, but what “ethics” entails, who gets to define it, and what kind of impact those questions have in practice \cite{green_contestation_2021}. Wong and colleagues \cite{wong2023seeing} note that ethics toolkits often lack guidance around how to navigate labor, organizational, and institutional power dynamics as they relate to technical work. Currently, the gap between principles and practice requires a lot of invisible work from technologists to navigate hierarchies, market dynamics, and the lack of awareness in the rest of the organization, which mostly goes unacknowledged \cite{deng_investigating_2023, wang2023designing}. 

Demands and idealistic notions of ethical reflection and action then clash with the reality of everyday industry practice, in which tech logics and organisational structures stand in stark opposition to the necessary resources required for ethical reflection - such as uncertainty, criticality, and slowness \cite{stark2009creative}. Yet little research has explored the impact and affective experience of integrating ethics tools into environments where such clashes are inevitable. We follow scholars of ethics in practice \cite{drage2024engineers, powell2022addressing, raji2021you, shklovski2023nodes} who argue against a too individualized approach to ethics, and consider ethics as a relational practice, where responsibility lies between the technical expert and the organizational structures within which they operate \cite{kranakis2004fixing}. We also build on research that argues for considering ethics as a situated and lived experience \cite{shilton2018engaging}, suggesting that the emotional context of ethics is important for bridging the gap between principles and practice \cite{dunbar2005emotional}. Prior research that reports on concrete methods, tools, and specific case studies of their usage, such as value dams \cite{Miller2007ValueDams}, and a variety of similar ethics interventions developed in HCI, tell us how practitioners deal with concrete trade offs and argue that those skills are transferable to future projects. They do not report on the longer term shifts in mindset and how that makes affected practitioners feel about their work moving forward. We therefore investigate ethical considerations as emotional and embodied \cite{su2021critical}, especially where they challenge existing social and political norms in an organisational context, coming into conflict with the techno-optimism that often accompanies technology development \cite{gould2009moving}.

\subsection{Emotional response, ethical sensitivity and moral stress}
Moral reasoning and ethical decision-making have an emotional component that has long been overlooked by responsible AI research efforts. Deep reflection that is required in design is a type of emotion work, that can come at the cost of feelings of guilt, self-blame, and emotional exhaustion \cite{ballam2019emotionwork}. Exploring ethical considerations in design through a soma design perspective, Garrett et al. \cite{garrett2023felt} and Popova et al. \cite{popova2022vulnerability} demonstrate that connecting ethical sensitivity to emotional experience can provide a generative approach to understanding the embodied perspective of ethical decision making, but also involves distancing and vulnerability as key emotional dimensions \cite{popova2023should}. Investigating an applied setting, Ma and colleagues report on the self-doubts, insecurities, and suffering faced by designers being limited in their decision-making power in addressing unfair business models through design decisions \cite{Ma2024SenseOfMorality}. Engaging with ethical considerations then, can be a challenging and vulnerable, emotion-laden process, but so far HCI has lacked the language to systematically describe the experience and its implications.

Outside of HCI, the challenges of moral conduct and ethics in practice have been well studied from a psychological perspective.
In fields such as management studies and nursing studies the fact that practitioners respond emotionally to moral quandaries and ethical dilemmas is established consensus. Studies from these fields provide us with defined vocabulary to identify internal processes around moral decision-making in practice. Jordan \cite{jordan2009social} for example defines the trait of having developed a situated attentiveness for the moral relevance of decisions and actions, in particular in connection to specific values, as \textit{moral awareness}. Based on a review of 108 studies of from fields such as medicine and accounting, Boyd and Shilton \cite{boyd2021adapting} review the concept of \textit{ethical sensitivity} for its relevance for technical practice. They conceptualise it as a three step process comprised of recognition (noticing an ethical problem, due to moral awareness), particularisation (understanding the situation that creates it) and judgment (being able to decide what to do about it).

Moral awareness and ethical sensitivity describe important parts of moral reasoning that precede morally motivated action and provide vocabulary to analyse the conduct and experiences of practitioners grappling with ethical challenges. An increase in ethical sensitivity and moral awareness is a common goal for practical ethics interventions in tech \cite{Boyd_2022_DesigningUp}. The assumption is that expanding the worldview of developers and designers will lead to engagement in deeper ethical reflection, which can result in increased ethical sensitivity, in turn leading to morally sound development efforts and responsible technology design \cite{vanhee2022ethical}.

However, increased ethical sensitivity and moral awareness can also have negative consequences \cite{weaver2007ethical}. An increase of self-reflective behaviour can make people more self-conscious and uncertain about their own actions, and a higher moral awareness is linked to higher levels of stress when roles and environment do not align \cite{Ames2020Antecedentsandconsequences}. Research from nursing studies reports that nurses suffer when the constraints of nursing work, such as the lack of time, supervisory reluctance or institutional policy, limit or prevent capacity for moral action\cite{corley2001development}. In management studies Reynolds and colleagues \cite{reynolds2012moral} identify suffering as a result of having to make morally relevant decisions, especially when there is a discrepancy between individual and organizational goals or views of the right actions. The discrepancy can come from having to choose between what are seen as right and wrong actions, but it can also occur when choosing between two legitimate but contradictory right options or two that appear equally wrong \cite{reynolds2012moral}.  

These adverse experiences as outcomes of increased ethical sensitivity are summarised by the concept of \textit{moral stress} \cite{reynolds2015recognition} (in nursing studies termed moral distress \cite{morley2019moraldistress, Lutzen2012Moraldistress, imbulana2021interventions}). Reynolds and colleagues define moral stress as "a psychological state (both cognitive and emotional) marked by anxiety and unrest due to an individual’s uncertainty about his or her ability to fulﬁll relevant moral obligations." \cite{reynolds2012moral}.
This type of stress arises from having to manage a mismatch between moral ideals and reality \cite{reynolds2012moral,reynolds2015recognition}. Emotionally, moral stress manifests outwardly through anxiety, frustration, and anger \cite{Lutzen2012Moraldistress}.
While there is disagreement about the precise conditions that cause moral stress, work load, time pressure, and role ambiguity are cited as the most common antecedents \cite{imbulana2021interventions,morley2019moraldistress}. When unaddressed, moral stress is reported to lower the quality of patient care or management decision-making, resulting in avoidance of morally relevant situations and even leading people to change workplace or occupation \cite{morley2019moraldistress}. Research suggests a range of organisational mitigation strategies in order to avoid cognitive and emotional burn out \cite{reynolds2015recognition}.

Despite the richness of research looking into operationalising ethical considerations into technical practice tools, few studies have explored the \textit{lived experience} of such operationalisation efforts. While the importance of emotions in ethical decision making is starting to be explored in HCI research, no conceptual bridge yet exists between operationalised ethics efforts and the affective experiences they trigger, especially where such experiences are negative. Ethics toolkits and similar materials to operationalize ethics are often expected to function like any other organizational process, while the surrounding power structures and required relational labor are overlooked \cite{wong2023seeing}. In addition, where ethical sensitivity has been researched in the field of technology design, it has been as a property of individuals, while technology design typically happens in teams and collaborative structures \cite{boyd2021adapting}. 

We present insights from an ethnographic study of technology development processes that have integrated ethics tools into everyday practice, paying attention to the embodied, situated, and emotional context of doing ethics with such tools. Leveraging the extant research on moral stress, we propose it as a valuable concept to provide vocabulary to help interpret our observations of ethical deliberation beyond acts of reflection, towards understanding the lived affective experience of practitioners. We see moral stress as an additional dimension contributing to our understanding of why well-intended ethics efforts struggle with sustainable adoption within industry.

\section{Research Context}
The effects of rapid digitalisation of the public sector launched a debate in Europe about the implications and ethical challenges of smart data-driven technologies in general and of their impact on the public domain in particular \cite{thylstrup2019politics, floridi2015onlife}. Multiple major European cities have responded by developing codes of conduct, principles, value sets, and manifestos to inform and guide the future development of data driven systems. The London Office of Technology and Innovation, a coalition of the GLA, the London Councils and the London borough councils, for example, established the London Data Ethics Service \cite{LOTI}, with in-house experts providing project facilitation, organisational development, and pan-London policy and research. Amsterdam developed the prominent Tada Manifesto \cite{TADA}, in collaboration with citizens, businesses, and governmental bodies \cite{AnAccessibleCity}. Barcelona launched the Ethical Digital Standards initiative \cite{EthicalDigitalStandardsToolkit}, which includes digital service standards, a Technology Code of Conduct, a set of methodologies for agile development in line with these standards, and the Ethical Digital Standard Manifesto \cite{EthicalDigitalStandardsManifesto}, which outlines the cities vision of technological sovereignty and digital rights for cities. 

Our study emerged out of a rare opportunity to observe the integration of such ethics tools into technical practice within a complex public services organization. We conducted an ethnographic investigation in collaboration with a European city that has launched their own responsible data manifesto and invested major efforts over the last five years to align their innovation agenda with this manifesto. The municipal organization brought together a team to develop their own version of a workshop format and an accompanying toolkit for it and to help integrate these tools into city operations. To ensure anonymity for our participants, we will refer to the city as the City, and will only describe the framework structure to provide context for our observations. In order to give the reader context, we will explain how we encountered and understood the ethics toolkit and its components. Rather than evaluating the ethics toolkit itself, we focus on the teams reactions to and experiences of the process of integrating the toolkit, as well as the social, emotional and organisational effects that the adoption of the responsible data manifesto had. We will introduce the team\footnote{For the sake of a more personal read, we have given the team members names, however, all names used in this paper are fictional.} and their roles, before we discuss their experiences with integrating the toolkit into their work.

\subsection{The Manifesto and The Toolkit}
In response to a deep concern about the impact of technology on the world, the City chose the form of a manifesto to express and communicate their principles. More than a code of conduct or a set of principles, a manifesto calls for collaborative and pro-active claiming of responsibility \cite{fritsch2018calling}. In addition to providing guidance, manifestos are meant to be persuasive, to demand attention, and to mobilize \cite{lyon1999manifestoes}. The City Manifesto names the concerns and uncertainties as “the big questions of the digital revolution”, calls for the forming of a community, and assigns responsibility to this community in order to make the City a moral thought leader and an example for how to address these big questions well. The Manifesto was developed in a broad collaboration with citizens, businesses, and governmental bodies, and provides a list of principles for the responsible use of data in the development of data driven technologies. It is accompanied by a wide variety of materials for distribution, such as posters, a website, and other promotional material. It signals the commitment of the City to addressing the challenges of developing smart city technologies responsibly and democratically. The City and involved parties consider the Manifesto a successful result of a complex, collaborative, and inclusive process. Over the last five years, the City has worked on implementing the Manifesto into their internal technology development processes, producing multiple initiatives to address and fulfil the expressed values. Recognising that the translation of generally formulated principles into concrete and practical action demands resources such as time, skill, and effort, the City committed to developing the required materials and competences internally. 

The work done by the City to follow through on their commitment to the principles of the Manifesto is a positive example of a public organisation working to close the gap between theory and practice in tech ethics. Over the last five years, through multiple initiatives, the City brought together an in-house team of ethics experts and designers to create materials and support that enable teams to translate the principles of the Manifesto into their work practices. One of the outcomes is a Toolkit developed by the team of in-house ethics experts and designers, specifically tailored to the work structure of the City's technical teams. The Toolkit consists of workshop materials designed to help teams develop a sensitivity for how ethical values are connected to their own everyday work and is currently slowly being adopted by project teams throughout the organisation. 

\subsection{The Toolkit workshops}
At the beginning of our fieldwork with the City, we met the group involved in the creation and management of the ethics Toolkit. Tanya was one of the ethics experts who was deeply involved in the development of the Manifesto prior to joining the City. Now, she is part of the City as an ethics expert with the mandate to help integrate the Manifesto into the City operations through the Toolkit. The work of integrating the principles into the City DNA, as Tanya frames it, has been a slow but steady process. Tanya's team has created videos and materials for employees to access, but the most effective way to become familiar with the Manifesto is to book a training workshop with Tanya or one of the expert facilitators. In our first in-person meeting at the beginning of the fieldwork, Tanya explains the intentions behind the workshop: 
\begin{quote}
\textit{“We worked out methods so that teams, when they get the time to think about ethical implications, they can do this in a structured manner, that it's replicable for several teams and projects, so they could kind of use the same method [in different projects]. And the results are shareable, they begin to help build this body of knowledge other people can learn from, that can be shared outside the organization [...]. And, especially, how to make sure that the theoretical idea of thinking about values becomes something that's workable, and something that seems [...] like they enjoy working [with it and feel like it] improves their work.”}
\end{quote}
The training workshop that Tanya's team offers is modular. When a team books a training workshop with an ethics expert for the first time, the first part is a general introduction to considering ethical implications when working with data. Tanya explains that she often has to start with the basics, as some people have no awareness of ethical issues at all within their work. This part is often tailored to the specific needs of the teams. After this introduction, the teams go through exercises that dive into the ethical implications of their specific project. Tanya says she considers it important to separate the introduction and the project-related part, because she wants to put the teams in charge when working through the project-related exercises of the workshop.

The materials for the project-related exercises include a huge poster, that spreads out like a tablecloth, putting the necessary conversations visibly on the table. The workshop format consists of four steps:
\begin{enumerate}
    \item Find the "North Star" asks the teams to specify what their project is about, and for whom. Tanya reports that at this step, teams often encounter unexpected results, even when they have been working together for a long time already.  
    \item The team goes through questions that guide them through the process of connecting the Manifesto values to their project. 
    \item The teams rate themselves on how they are doing on each value, and list the risks associated with lower scores in a diagram. 
    \item After the teams have the risks of their project listed in the diagram, they brainstorm ideas on how to mitigate them, and then prioritize mitigation strategies. The teams formulate the outcomes of this process as a project brief for themselves. The brief is the final part of the big poster, and is designed to be cut off, so that the team can hang it up at their desks and keep handy for checking again later. 
\end{enumerate}

After the training workshop, the expectation is that the teams will run their own smaller version of the workshop for each new project and revisit the methods and questions throughout the projects. The materials for each of the four steps are always available to the teams online so that they can repeat the exercises as desired. 
Tanya recommends that at the end of the training workshop, the teams choose a one member to take on the role of the internal \textit{Ethics Owner} \cite{metcalf2019owning}. \footnote{The team uses a different term to describe this role, however for annonymisation purposes we use the term Ethics Owner, defined by \cite{metcalf2019owning}, as it is similar in meaning.}, who is in charge of ensuring the ongoing engagement with the principles, either by setting up and facilitating recurring workshops themselves or by delegating it to other team members.

\section{Data collection and analysis}
We were invited to spend time with technical teams at the innovation department within city operations in the fall of 2023, when the manifesto had been around for 5 years, and the developed workshop materials had been used for about a year. The first author worked with two teams for 5 weeks in person and attended meetings and sessions online for a total of three months before and after the in-person period. The teams we collaborated with work on data driven smart city projects that provide services for both civil servants and citizens. The teams focused on small scale implementations of technical ideas, building prototypes, proof of concept applications, and "first versions" of smart-city technologies. These are then used by decision-makers to discern how ideas could be turned into reality, what kind of opportunities and risks that would bring, and what kind of technical, architectural and user requirements would be necessary to make it work. In this way, both teams provide research, ideation, and prototyping, as well as technical development of smaller scale projects. Following the release of ChatGPT, AI has received increasing attention within the municipality and technical teams working with ML technologies are facing lots of interest from other departments. The team under study in particular considered themselves responsible for showcasing what digital technologies and AI can do for the city. 

The first author leveraged their prior expertise as a designer to collaborate with the teams as an embedded researcher on one of their exploratory projects - the development of a routeplanner for people with mobility impairments, that aims to create personalised and targeted routes depending on accessibility needs. The tasks during this project included user research with wheelchair users, design sketches, development of technical requirements, data infrastructure and technical prototypes for route testing. 

This project only provided the background for the first author to be integrated into the team. As such, research focused on team tasks and operations more generally during the experience of using ethics tools for several different projects. The first author was able to observe how the team members talk about and perceive their work and its ethical dimensions, as well as how they integrate the practical tasks into their everyday work, negotiate required resources and requirements, and schedule, scope and define resulting work tasks. This allowed the first author to experience how each team defines itself in relation to the rest of the organization, how they see their internal and external responsibilities, and their team purpose. In addition to mapping the logistical challenges, we paid attention to the feelings and emotional layers emerging through reports, team communications and actions. The observed work projects provide illustrative examples for how the affective experience around ethical decision making unfolds. In our analysis, we focus on the positive and negative emotions connected to the processes, rather than the outcomes of the processes by themselves. We therefore focus in this paper on the team that continuously worked on integrating the toolkit into their work practices in order to demonstrate the challenges for an engaged and invested team of practitioners.

The teams worked on up to 3 projects at the same time, thus allowing the first author to observe and learn about the daily operations across multiple project contexts. In addition to participating in everyday work life as a temporary part of a project team, the first author conducted interviews with all team members, attended regular team sessions such as stand ups, sprint planning, team retrospectives and task refinement sessions, and joined additional team sessions scheduled as needed. Interviews and meetings covered a broad range of past and ongoing projects in the City. The first author also attended joint meetings between the teams and other stakeholders, and interviewed other employees of the city, such as managers or close collaborators. In total, we conducted 13 formal interviews, attended 30 team meetings, and 15 other non-team specific meeting sessions. Interviews were recorded, and meetings were documented through extensive note taking. Throughout the stay, a diary was kept for daily note taking and reflections to document daily observations.

The research team maintained a weekly correspondence where the first author sent regular ethnographic reports to the rest of the team during the fieldwork, engaging in joint discussion and reflection over email to facilitate ethnographic theorization \cite{cerwonka2008improvising}. After the fieldwork, the interviews were transcribed and open coded alongside ethnographic notes, followed by thematic analysis by the first author. Emotions and affective experiences crystallized as recurring and relevant to the teams experience of ethics interventions. During the analysis phase, the research team recognized signs of moral stress in the data and used it as a sensitizing concept to explore the language used to express emotional experiences and relational practices. We then synthesized the findings into their presented format. Study participants gave comments on early drafts of this paper to ensure consistency and correct understanding of city operations.


\subsection{The Team}
When the first author began fieldwork, the team under study had already run a training workshop with Tanya for the route-planner project, and had been familiar with the Manifesto for about 6 months. As an outcome of the training workshop, the team had selected one of their members to take the role of the \textit{Ethics Owner} to guide their engagement with the manifesto going forward. To provide context for our findings, we first introduce the 5 core members of the team:

\begin{itemize}
    \item \textbf{Sarah} is an AI researcher and has been with the team for 5 years. She has taken on the team lead position in the last year. 
    \item \textbf{Paul} is an AI specialist and has joined the team 3 years ago. He has always been interested in inclusion and diversity, pushing for more accessibility projects. 
    \item \textbf{Nadine} has been with the team as data scientist for 4 years, and has taken on the role as the team Ethics Owner. 
    \item \textbf{Eli} is a junior data analyst. She was invited to join the team a little over a year ago by Sarah, after doing her bachelor thesis project at the municipality with some people from the team. 
    \item \textbf{Johan} is the project manager in the team. He has worked at the municipality as a freelancer for a few years and has joined the team officially about a year ago. 
\end{itemize}

After the training workshop with Tanya, the team was prepared to run their own workshop sessions internally. Nadine has copied the templates for the workshop onto a collaborative online canvas. For each new project, she copies a new template next to the existing ones, so that all of the documentation for previous workshops is in one place and visible during future sessions. Before our visit, the team had run one more workshop by themselves for an ongoing project. During fieldwork, the first author attended one follow up workshop for the route-planner project, and two workshops for newly started or planned projects. 

\section{Findings}
Researchers have argued that for ethics interventions to be successful there must be a combination of tools and practices developed with existing work structures and practical requirements in mind \cite{madaio_co-designing_2020} as well as emotionally invested and open-minded practitioners willing to engage and reflect \cite{frauenberger2017action, metcalf2019owning}. The City has put a lot of effort into creating the requisite tools. As the workshop becomes slowly adopted across the organisation, Tanya reports on the experiences the participants face: 

\begin{quote}
\textit{“[The Workshops] create a space where people that maybe have doubts, somewhere in the back of their minds... it helps you bring out some of the uneasiness, maybe something feels off but doesn't really have a space, like in what meeting are you going to address this lingering tiny thing in your mind? And the workshop was really designed to give everybody a voice to put that on the table. Which can be hard, or tricky. But people do get to interesting insights, I think. Quite often you find that there's tensions in the team, that people do have all kinds of issues with what they're doing, they have their doubts, and then that all comes up during the session.”}
\end{quote}

As Tanya mentions, the workshops give the teams the space to surface lingering ethical concerns and to sharpen their moral awareness around the work challenges they face. The training workshops offer an introduction to the general ethics initiative across the organisation, but they are only the kick off for an expected ongoing process for the teams. Teams often leave the workshop with more questions and open disagreements than they were aware of before. In order to settle these, they need to continue to develop their ethical sensitivity, internal vocabulary, and communications around the ethical implications of their projects. It is at this point, where overwhelmed teams might abandon the process and stop engaging with the toolkit moving forward. The team under study however was ethically engaged and committed to continue figuring out new questions and emerging doubts, as evident from their own efforts in repeating the workshops for new projects and their openness to our fieldwork engagement. Multiple team members indicated that they were interested in hearing an external perspective on their efforts, because on some level they felt that perhaps they were doing something wrong in their efforts with the Toolkit. During the very first week, Paul states:

\begin{quote}\textit{“I am really interested in hearing what you think after spending time with us! You can tell us what we are doing wrong!”}. Johan, the project manager, is diligent about making sure that we report back to them after the visit, so that they can learn from our insights and observations, and Sarah concurs: \textit{“I mean you saw some of our difficulties we have... It would be great to hear what we can improve. I am sure there is so much we are not doing great at the moment.” }
\end{quote}

The team's reactions to our visit show that they are committed to ensuring that their work results in ethical outcomes (do the right thing) and that they use the provided materials well to achieve those (do the thing right). Throughout the visit, the team expressed eagerness and anxiety, reaching out to an external expert to be helped and reassured, to \textit{tell them what they are doing wrong}. In this way, they were giving voice to the moral stress they experienced due to the newly developed ethical sensitivity, which expressed itself as an underlying frustration and ongoing struggle to find guidance and avoid uncertainty \cite{Lutzen2012Moraldistress}. As will become visible in the following sections, the conversations where the team reflected upon and questioned their work practices are noticeably laced through with affective language, making visible the moral stress that accompanies moral awareness and is coupled to the ethical tensions being discussed through the ethics workshops. These simmering emotional responses to the ethics efforts caught our attention. The team embraced the questions and ethical implications raised by the Toolkit, and did their best to respond to them. Literature suggests that moral stress is an important concern when developing ethical sensitivity in organisational contexts, especially where such interventions are expected to deliver satisfying results \cite{reynolds2015recognition}. In the following section, we trace the duality of eagerness and insecurity the team displayed across their work processes. We identify where moral stress comes up in different forms, how it is articulated and navigated by the team, and how it affects them beyond the logistical demands of navigating trade-offs.

\subsection{Relating ethics across organisational structures}
Organisational structures play a big role in how production teams perceive their capabilities and power to address ethical implications of their work, when they become aware of them \cite{widder2023dislocated, wong2023seeing, madaio_co-designing_2020}. The team in our study faced a number of challenges, though partially alleviated through the proceduralised nature of the Toolkit, that provided space and a vocabulary to express ethical sensitivity in regards to their work. This organisational effort gives space for other tensions to become visible - tensions and stress that originate from relational friction, but are ultimately experienced individually \cite{reynolds2012moral}. Organisational structures, such as management hierarchies and project schedules, are a necessary part of managing technical projects. However, they not only structure work processes, they also structure how people across and beyond an organisation relate to each other - they create relational configurations, within which different facets of moral stress unfold. In order to understand antecedents and expressions of moral stress we attend to the relational configurations that the team navigates with and through stressful emotions. Other researchers \cite{pillai2022exploring, su2021critical} have highlighted the impact that relational factors play in how designers and engineers react and deal with moral pressures. Pillai and colleagues \cite{pillai2022exploring} for example used the differentiation between micro, meso, and macro layers of closeness. We find a similar importance of distance and closeness within the relational configurations but ground these in organizational structure: relations within the team, relations beyond team within organization, and engagements with stakeholders beyond the organization. 



\subsubsection{Within Team: Owning Ethics}
The team itself is the first space where tech workers explore and potentially confront ethical dilemmas of their work. After the initial training workshop the team itself provides the closest social frame for negotiation, confrontation, and decision making around ethical implications of the projects. As Ethics Owner, a large part of that ongoing process is driven by Nadine, who schedules and sets up ethics workshops for different projects, and takes care of integrating them into the existing processes. One example of this work is her task to transfer ideas from ethics workshops to the team's backlog. Nadine aims to create explicit ties between the principles the team discuss during the workshops and concrete tasks that she can put on the backlog for the next team sprint. She considers this the best way to integrate the responsibility for ethical implications as shared, \textit{“normal”} work for the entire team that is integrated with other work tasks, rather than treating them as special efforts. She does this by re-framing the ideas of the workshop as ticket items that are collected in the backlog of the team - an ongoing to do list, where the team collects outstanding tasks that will be taking into consideration for planning the future allocation of time and resources. The team works in three week intervals, called sprints, and at the beginning of each sprint, the team visits the backlogs and plans which tasks from the backlog to pick up in the coming three weeks, who should work on it, and what constitutes successful task completion. Nadine explains: 
\begin{quote}
\textit{“the goal is to get the stuff [the ideas from the ethics workshop] we want to do on the backlog.[...] And anything of reasonable size that we work on has to be here otherwise it will not be considered to be picked up. So our only way to get this stuff [she points at the post its with ideas from the workshop] done is to get it in here [she gestures at the backlog]. Yeah, and the only thing I need to think about is.. I need to think a little bit about formulation. And I need to think a little bit about like, what's the first step."} 
\end{quote}
This process of transferring ideas originating in the ethics workshops into the ongoing work process is the doing of ethics, where Nadine takes the responsibility of turning a discussion into a first step that is actionable and feels familiar enough for the team to act on it. She sees this as the only way to ensure that implications discussed in the workshops get addressed. The Toolkit provides a vocabulary and a frame for how to engage and apply ethical sensitivity. As a follow up the team finds touch points between the other things that are going on and their insights from workshops. Nadine channels a constant stream of translation and integration between reflections, insights, and work tasks through her role by turning ethical inquiry into more mundane tasks that can survive in the standard formats of project management. Without this translation work of the ethics owner, the ideas from the workshops only exist as good intentions, without a concrete tether to the work reality of the team. From a process point of view, this translation work should act as a mitigation strategy, in that it transitions ideas with moral relevance into technical tasks that can be fulfilled without further reflection. In practice however the task of this translation work focuses moral stress on one person to make moral decisions for the team, and to provide mitigation for the rest of the team members \cite{Lutzen2012Moraldistress}.

Nadine is able to perform this translation work because she has both internalised the ethical commitments of the Manifesto, and is an integral part of the team and its culture of collaboration. As Tanya comments, the teams are in charge of making the actual judgements about which costs they are willing to take on to address the implications uncovered in the workshops. To ensure that ideas evolve into tasks that are narrow enough to be acted upon, Nadine needs to apply her personal judgement to what she considers necessary for a task to be performed in accordance with the ethos of the ideas that spawned it. The moral stress emerges exactly at the seam between reflection and actions, where toolkits and processes are placed to enable the transition \cite{imbulana2021interventions}. For example, one idea from the workshop is to define a second type of user persona, that has more complicated accessibility needs, and that challenges the team to expand what they consider accessible. While documenting this idea as a ticket item in the backlog, she adds "collaboration with a designer" to the task description:
\begin{quote}
\textit{“That is something I took the freedom to add myself. So I think in general, it's just not nice to have either the person working with the users or the project leader, not included in this kind of large decision. I think it kind of… it will shape the project quite a bit. I would be worried that somebody would forget it so I took the liberty to write it down. [...] And now, what I'm considering is if I want to include in this ticket also like investigating this user persona and find out what is needed or if I want to do that separately... I think I want to do it as one because otherwise things are just moving so slowly."} 
\end{quote}
In addition to formulating the first step of turning an idea into action, Nadine applies personal judgements to modify and amend task formulations to ensure that they deliver on their intention. Through these judgements she bridges between the team work practices and the intentions of the ethics workshops. In this case she is aware that the definition of the task, who is included in it and who is not, will have a big impact on how the project turns out, and takes individual measures to adjust task instructions accordingly to comply with the context within which they will be executed. This is a sophisticated interpretation and understanding of the influence of relational set up of tasks that she applies, both in regards to the construction of the task, how it will be received by the team and how it will impact the future of the project. This role of the ethics owner is not only an additional work load to her normal tasks, but one that is explicitly morally charged. While she is personally invested in the quality of her work, and cares about ethical implications a lot, Nadine describes the role as stressful:
\begin{quote}
\textit{“I really don’t like the term Ethics Owner. It feels heavy. I am the kind of person when I have responsibility, I take it very seriously. So if you call me Ethics Owner, it is just too much on my shoulder.”} 
\end{quote}
The heaviness that Nadine notes is a by-product of what ends up being an individualized responsibility for morality \cite{Lutzen2012Moraldistress} that she takes on in the role of the Ethics Owner. The ideas build pressure, they construct the \textit{shoulds} that the team feel hanging above their heads.


While Nadine takes on the responsibility of facilitating the translation from the ethics workshops to the backlog, the team supports her efforts and contributes to sharing the responsibilities by being responsive and inquisitive during follow-up meetings, such as sprint reviews and task refinement, where they plan and evaluate their project work together. Overall, this strategy seems successful, and others in the team explain that they appreciate the way Nadine translates the abstract discussions during the workshops into concrete tasks. In some cases the team calls each other out when they plan too optimistically with their work time, or are too loose in their definition of what defines a task as done. So while Nadine carries a deciding role in the constant stream of reflections and project planning, the team has also integrated this role into the social support network they provide each other, and thus provide mitigating structures by dispensing stress through social support \cite{imbulana2021interventions}: 

\begin{quote}
\textit{"I think we actually depend quite a lot on reviewing each other's work. It's pretty standard for most of the things we do, that somebody watches along and we expect feedback. And yeah, I actually feel this is going pretty well. So I think, just this pointing out that something is not completed can be uncomfortable and people don't always like it. But more generally, like the reviewing of each other's work, I think that's going pretty well and it's kind of well appreciated by my team. [...] And what I do enjoy is that there's a lot of freedom. I think people are kind. They think really deeply about subjects. [...] there's also quite a lot of room now to debate on these things."} Nadine reflects upon the teams interactions.
\end{quote} 

In this way, the team applies the same care to the additional tasks resulting from the ethics workshops as with any other part of the project, and support each other in navigating the emerging spaces of uncertainty \cite{shklovski2023nodes}. The moral stress that arises from the ethical sensitivity that Nadine applies during the translation of ideas into tasks is, if not absorbed, to some extent alleviated through an emotionally responsive team culture. The stable social relations of the team provide an emotional buffer and space for moral stress to be addressed and mitigated through shared recognition \cite{imbulana2021interventions}. Yet throughout the fieldwork team members repeatedly expressed frustration that integrating ethics insights continued to be challenging, convinced that they must be doing something wrong.

\subsubsection{Beyond team, within organisation: The tech hype and fear of missing out}
This section describes the relations of the team as an entity with a specific identity within the organisation, with respect to the internal stakeholders they communicate with. Conflicting obligations within an organisational context can often cause stress and moral stress in particular emerges when moral obligations are in conflict with organisational roles or mandates \cite{reynolds2012moral}. The team is embedded in the innovation department of the municipality, part of a broader organization where they collaborate and communicate with a variety of other teams and stakeholders. The release of ChatGPT has led to a steep rise in interest from other sections in the department, who are curious about using AI in their work processes to make their work easier or more efficient. Nadine describes the current position of the team in the organization where the municipality is trying to find their direction around the use or non-use of AI technologies: 

\begin{quote}
\textit{“One of the challenges I do think is the tech hype, and that there is like, oh, there's this interesting new technology, we have to do something with it. So people are sort of expecting something, but really not clear about what they're expecting. So you're in some weird middle ground where you have on one hand really a lot of freedom, but at the same time, not that much freedom because so many people are watching you.”} 
\end{quote}
The team is under pressure to deliver on the promises of this exciting new technology, that currently feels like an under-explored opportunity for a digitally progressive city. Just as many companies follow the streams of innovation-culture hype, the municipality lives under the pressure to keep up with the technical developments pushed by media. At the same time, the Toolkit workshops add additional demands on the team to be conscious and reflexive about the motivations and intentions behind their technical work:
\begin{quote}
\textit{“It seems that even though there's all this fear mongering, I think when people want to use AI for their work, they just want the benefits and they don't want to think about it right,”} states Sarah. 
\end{quote}
One of the requests is an internal version of ChatGPT for the municipality, which upper management wants to develop “\textit{so we can put that out there and make informed expert recommendations}”. However the team keeps running into confusion about the project's goals and intentions. The question of \textit{“who are we doing this for and why are we building this?”} comes up in multiple conversations, such as the ethics workshop, the backlog refinement, and the sprint planning. During the ethics workshop for another, similar AI project, the question for the motivation to use LLMs comes up again and Paul wonders: 
\begin{quote}
\textit{“Why do we do this? Are we now automating something that was supposed to be human effort anyway? Or, you know - why is this the solution to the problem? [...] Is it just because LLM is a convenient hammer that makes every problem look like a nail?”}
\end{quote}
This expression of friction between the push for innovation-driven fast development and considerate, well informed project decisions arises from the fact that the workshops add more demands and pressure on teams, in the context of a culture that is already charged with time pressures to deliver technical innovations at fast speed. The willingness of the team to reflect beyond the technical feasibilities of their potential projects and the underlying intentions of their tech ideas confronts them with the limitations of their own roles and responsibilities. As the technical team, they are the experts that the municipality comes to for consultations about feasibility and testing, but the introduction of the Manifesto and the Toolkit also asks of them to follow their ethical sensitivity to an extent that is not mirrored throughout all the layers of the organisation. Organisational structures provide a structured separation of concerns \cite{Subramonyam2022Seperations}, to enable clear delineations of responsibilities. However, the Toolkit and the training of technical teams in ethical sensitivity result in loosening of formal and informal separation of concerns, leading to the team being confused and unsure about where to draw the line around their responsibilities. 

The City has legal teams that are seen as responsible for ethics-related decisions especially in the context of compliance with regulation. Yet the workshops teach the technical teams to recognise ethical implications that go beyond questions of legal concerns, compliance and regulations: 
\begin{quote}
\textit{“The less clear questions are, of course, the more vague ethical feelings that you might have about something where, you know, you could somehow frame it in a way that meets all the official regulations, but still, your gut feeling says that it's not a great idea.”} Johan says. 
\end{quote}
In these situations, where the municipality is still working on finding direction, the team works to balance their responsibilities as civil servants as well as technical experts. This puts the team into an unofficial role as gate keepers, with the implied responsibility to judge which risks are ethically relevant enough to raise for discussion beyond the team and which are not. As informal gate keepers,  they feel responsible to consider the ethical consequences of potential projects, without necessarily having the mandate to change them – something generally associated with legal teams, who act as gatekeepers on legislative grounds: 

\begin{quote}
\textit{“In the end, it's going to be a sort of political choice that a manager at some point needs to sign off on.”} Johan shares. 
\end{quote}

The term "political choice" has a rather specific meaning in a municipal context, highlighting tensions between capital P Politics (as in political programs developed by parties) and small p politics (as in implicit value judgements about who and what to prioritise) which civil servants must navigate within formal governance structures \cite{Boehner2016DataDesignandCivics}. Capital P "Political Choices" can only be made if there is explicit mandate to do so, and such choices get escalated back up to the political decision layer. "Political Choices" are by definition decisions that civil servants cannot make on their own. Tensions arise in differentiating when a dilemma is a managerial (political) choice, and when it is a Political Choice. 

While the team is supported through the framing of the tools to open conversations amongst each other and with other teams and management, the identified way forward is grounded in political decisions that are often out of the team’s locus of control. The tools give the team the capacity and space to reflect on the ethical dimensions of their work, but only limited power over the consequences of such reflection and mandate to address them \cite{seberger2021empowering}. The team acknowledge their limits and limitations and generally aim to operate within the boundaries of what they can deliver. However, in contrast to the internal team dynamics, the structures in the organisation do not provide the relational resources to navigate these limitations in ways that offer the team assurances about their insecurities. The moral stress that results from the dissolving boundaries around separation of concerns \cite{Subramonyam2022Seperations} triggered by the workshops has nowhere to go, remaining with the team to deal with. 

\subsubsection{Beyond organisation: good intentions and the status quo}
The municipality is part of the broader City governance configuration together with citizens, businesses, and other organizations and the team has to grapple with the tensions that arise between their own projects and intentions and the outcomes for the citizens they are serving.

The Manifesto postulates high-level principles and values in relation to citizens of the City, for example stating that data usage should benefit everybody, that models should be transparent for everybody to understand, and the like. The Toolkit and workshops have taught the team to think beyond the technical aspects of their work and consider the broader social and political impacts their projects might have. But beyond these reflections, the workshops also build pressure to act in accordance with those insights and construct strong ideals for the team to live up to, with a long list of "shoulds" hanging over them. So while the workshops give the team the tools to \textit{decide} on ways forward, the actual discomfort around getting these tasks done often surfaces outside of the workshops, in particular when it comes to actions that relate to stakeholders outside of the organization.

For example, the team identified an ethical dilemma in a computer vision project. In an attempt to gather more information about the accessibility of the city public infrastructure, the team worked on a model to help measure the real width of sidewalks. While sidewalks are planned with a certain width by city planners, this is later compromised by additions of permanent fixtures such as polls, trees or signs, or temporary obstacles such as roadworks, wrongly parked vehicles or trash. The project was intended to enable support for people with mobility impairments who rely on wheelchairs or other mobility assistance mechanisms and require a particular width to manoeuvrer. Having developed the model however, the team realised that the tool they created also brought up a number of questions about surveillance. Aside from estimating sidewalk widths, the algorithm could pick up on citizen digressions such as mis-parked bikes and misplaced flower pots, \revision{which could be penalized more efficiently without actually helping wheelchair users. While trying to create a system that would provide one group with more useful information about the city infrastructure, the team realized that they unintentionally created a system that could also reinforce a punitive approach in cases irrelevant to their purpose (i.e. where large flowerpots do not in fact intervene with accessibility) or lead to punishment without any resolution (such as creating fines for misparked bikes without providing sufficient space for both bikes and an accessible sidewalk), something the team was not comfortable with supporting.} Reflecting on this project in particular, Paul worried: 

\begin{quote}
\textit{“We were so well meaning, so well-intentioned, and afterwards I thought, oh, should we even do this?” }
\end{quote}

Of course, just because an algorithm can be used in such a way doesn't mean it will be. The team put extra effort into ensuring that potential for misuse is recognized, so that it could be curbed proactively. However, given that the teams also want to be transparent about the models they develop and potentially share algorithms with other parties, such worries remain potent. As Sarah explained: 

\begin{quote}
\textit{“We, as a team, have the same responsibility [as all civil servants to raise issues and concerns] and maybe even a stronger responsibility when it comes to AI”.}
\end{quote}

After this project Eli has started to see the infrastructures of the city differently. She now notices bikes that are parked blocking the sidewalks and other obstacles in the streets that make the overall infrastructure less accessible. She has started to make sure that she does not park her bike in the way of a potential wheelchair path and she moves other bikes when she can, getting frustrated when she realizes that the existing infrastructure does not take all needs into account similarly. During one ethics workshop for the accessible route planner, she discusses with the others how to draw the attention of officials to the issues of lacking infrastructure and challenges for mobility impaired citizens. The group agrees that to match the ideals of the Manifesto, broader changes to infrastructure are needed, more than they can do with an app. \revision{On the one hand, the team feels responsible for providing technical support for a more accessible city. Considering the necessary changes beyond the production of an app also becomes important for their agenda. They therefore decide to keep the topic of infrastructural changes in their documentation and add an additional post-it to their idea board with the task of sharing data on obstacles with officials to advocate for more accessible infrastructure.} On the other hand, the team needs to be conscious about their own resources in order to get something done, even if it is not living up to the full scope of the ideal:

\begin{quote}
\textit{“We don’t have the capacity to immediately develop something that works for everybody. I think one thing we have to do here is clearly communicate what the scope of this pilot is, and make sure that it is also very clear for the public that we’ve consciously made this choice and are aware of the fact that we are not helping everybody.” Sarah states. }
\end{quote}

The next day, Nadine reflects on how building an app that supports people with mobility impairments might even relieve pressure off other departments to design the city more inclusively in the first place, which would be the opposite of what the team would like to achieve: 

\begin{quote}
\textit{“Like, you have discussions and for some things, you think okay, yeah, we could be a bit more transparent, that'd be cool. But this one was really hitting me in the face like, okay, I think we're doing … are we not really consolidating the situation instead of changing it?” }
\end{quote}

Deep reflections on far reaching consequences often spark more questions and frustrations when tech solutions treat symptoms rather than causes. In the case of the accessible route planner, any fundamental change in infrastructure is far outside of the mandate or jurisdiction of the team. The team struggles with the problem that the systems they encounter in their projects are built into an environment with already established structures that pull towards a status quo. This has two consequences for the team: first, these structures are not in line with the values in the project and thus can easily subvert it; second, the team can only make progress from the given starting point, but tends to judge itself from an ideal interpretation of values and responsibilities promoted by the Toolkit. In attempting to live up to the idealistic values the team began to recognize that sometimes tech solutions are just band-aides to bigger issues. All their efforts to improve the ethical scoring of their production process – by making it more inclusive and transparent – might not create the end result they would prefer, simply because their potential solutions are tools and not systemic changes. 

\section{Discussion} 
In our study, the overall effort and initiative of the municipality can be considered a success - not only did the technical team we studied adopted the suggested methods and engaged with them continuously, they also used the tools to push for project changes in line with ethical considerations. However, the findings also highlight that a success story of an ethically invested tech teams plays out differently from what is expected by toolkit designers, business managers, and even technical professionals themselves. The team experience of increased ethical sensitivity \textit{feels} very different from what ethics toolkits and workshops might promise, as they encounter new vulnerabilities and insecurities. Vulnerability is endemic to the process of accepting responsibility for ethical decisions \cite{ballam2019emotionwork}. This is not only a matter of distributing mandates within the organisational structures and creating appropriate processes, but also of accounting for what Popova and colleagues term intense somatic experience of discomfort \cite{popova2022vulnerability}. While recent work by Popova and colleagues \cite{popova2023should} notes the importance of emotions in engaging ethical issues, what they describe maps directly on expressions of moral stress and the most typical mitigation strategies of distancing and abstracting described in the moral stress literature \cite{imbulana2021interventions, reynolds2015recognition}. In the following section, we discuss what we see as transferable insights from our findings on the emergence of moral stress for HCI.

\subsection{Beyond reflection: Engineers are set up to fail and expected not to care} 
Reflection is often seen as a necessary component of ethics tools and methods \cite{devon_design_2004,frauenberger2017action} that can result in important insights \cite{markham2020taking}. Approaches such as VSD and other interventions developed within HCI have been successful at broadening perspectives, including other voices, and providing productive opportunities to foster the development of moral awareness and ethical sensitivity \cite{friedman2019value, chivukula2021surveying, wong2020beyond, ballard2019judgment, umbrello_mapping_2021, Miller2007ValueDams, winkler2021twenty}. What has been lacking is a discussion of the costs that these approaches create. As ethics workshops and toolkits teach technical teams to reflect and question, they inevitably create more questions than answers and thus destabilize previously held certainties. Rather than finding clarity, feeling more assured in the goodness of their work, and capable of delivering ethically sound projects, the team in our study experienced moral stress from uncertainty and lack of assurance as they identified and negotiated increasingly difficult trade-offs, which manifested as anxiety and frustration \cite{Lutzen2012Moraldistress}. Ethics toolkits construct the values they consider as ongoing sociopolitical practice \cite{wong2023seeing}. They also construct the people using them, in this case technical practitioners. Many ethics interventions not only consider technical practitioners as not yet ethically engaged, and missing ethical sensitivity in order to build technology responsibly, they also consider technical practitioners as disembodied, impartial actors, who can work through the expected process of reflection, identity work, and required action without much emotional reaction to shifts of perspective and awareness of positionality and responsibility. Our findings show that rather than an indicator of shortcomings or failing work processes, discomforts from considering ethical implications are indicators of internalising a required level of commitment and emotional investment. In other words, when engaging with any effort to train and increase ethical sensitivity, moral stress is always part of the deal, and an indication of growth rather than a failing \cite{reynolds2012moral}.

The affective experience of moral stress goes counter to the general narratives of what tools, pipelines, and work processes are expected to do in a technology innovation context, where creative friction is welcome as long as it does not impose on productivity \cite{stark2009creative}. While no ethics interventions claim to provide solutions to ethical dilemmas, they typically offer guidance or support in making decisions, thus implicitly promising to ease the burden and improve outcomes \cite{chivukula2021surveying, morley2020initial}. This approach originates from a culture of solutionist thinking, where problems are always coupled with solutions, even if the problems are inherently unsolvable \cite{cunningham2023grounds, sicart2020pataphysical}. When it comes to ethics interventions, such coupling creates expectations that dealing with and addressing ethical dilemmas through the provided exercises is a skill to be trained, becoming easier with time, and resulting in ‘better’ outcomes. 

These expectations fall apart when aspirational values and idealistic resolutions collide with real-life working conditions, and the interventions fail to provide unquestionable paths of action. Practitioners expected to engage in ethical deliberation during their technical work are automatically set up to fail on one front or the other, never able to fully reach the idealistic vision portrayed by value sets, manifestos or other prescriptive ethics materials. While VSD-informed methods such as value dams \cite{Miller2007ValueDams} offer practical and actionable approaches for addressing value tensions, such processes in themselves do not provide a lens to understand the emotional and affective impact their application has on the professionals adopting them. This is where we suggest moral stress as a beneficial lens - in addition to evaluating the ways in which such tools provide approaches for ethical dilemmas, we argue that, since ideal resolution is impossible, we also need to understand and account for the emotional fall-out of this `failing'. 
The moral stress of having to adjust personal perception of technical work as ethically productive is rarely acknowledged, leading to harsh experiences, such as ``being hit in the face'', as Paul puts it, with the realization that well-meant technical interventions might contribute to consolidating existing exclusionary structures. Feelings of vulnerability then are more than “an active ethical stance” as Popova and colleagues argue \cite{popova2022vulnerability}. Rather, they are endemic to making ethical considerations explicit in technical processes and practices and can manifest as moral stress. 

Where moral stress remains unacknowledged and hence unaddressed, ethics interventions - no matter how ostensibly effective and well designed - will struggle to achieve the hoped for changes. To mitigate the costs of emotional labour and moral stress, people may distance themselves from the points they should intentionally engage, undercutting the original intentions of the interventions \cite{wong2020beyond}.
According to prior research on moral stress, one common mitigation technique is limiting the view of the broader picture and narrowing the space of responsibility for the affected workers \cite{imbulana2021interventions}. This is in direct opposition to ethics interventions, which seek to increase ethical reflection and broaden the scope of relevant concern. Recognising the broader picture and the relevance of ones own work within it is important to the ideals of responsible computing but this can make it harder to see small wins rather than bigger failures. This is evident in the struggles of the team around their potential beneficial impact vs. supporting the status quo. Yet focusing on small wins can be an effective tactic working for broader changes in constraining environments, what Horgan and Dourish describe as aspects of tempered radicalism in technical practice \cite{Horgan2018-HORAAA-19}. This is where 'small p' politics and moral stress connect. In order to mitigate moral stress and be productive in 'small p' politics, limited views of impact are productive but in tension with the general ideal of developing broader moral awareness. This leads to the required next step that follows after the implementation of ethics interventions - from developing moral awareness towards managing moral stress as they negotiate the discomfort of ambivalence and disjunctures between the moral ideals and the constrained realities of technical practice in organizational contexts. Integrating such recognition into work practices demands emotional labour of the team members individually and collectively, in order to accommodate a shifting perception of their roles, tasks, and responsibilities. 


Here, the capacity to lean on a community and safe social relations, as productive coping mechanisms for difficult feelings stemming from ethical quandaries, becomes crucial \cite{pillai2022exploring}. Our findings align with prior research in nursing studies, that states that a socially healthy team culture is a non-negotiable pre-requisite for teams to productively engage in ethics efforts that continue to put them in difficult and uncomfortable situations \cite{imbulana2021interventions}. Well-functioning teams are not the domain of HCI, and there is a wealth of literature within organisational studies and management studies that address this. However, team relations must be a consideration when designing ethics interventions, because these inevitably rely on a solid relational foundation within teams and organisations, where critical and difficult topics are being surfaced. The Toolkit, as its creators pointed out to us, is designed to create spaces where doubts can be surfaced and openly interrogated. Such ''spaces of doubt'' are key to the doings of ethics, even as the ethics owner on the team works to translate these into ''nodes of certainty'' as concrete steps on the backlog \cite{shklovski2023nodes}. The team we observed in our study had an extremely supportive and kind culture of openness that enabled them to share the emotional costs of uncertainty and doubt when engaging with the ethical implications of their work. They appreciated this space for reflection but also clearly articulated the increased anxieties and frustrations that resulted in such engagements.

\subsection{Responsibilities of organizations: Stress that has nowhere to go}
Ethics toolkits are always a moral comment on the existing practice, putting teams on the spot to make normative decisions and take moral stances. As our findings show, the spaces for reflection created by the Workshops stand in contrast to a culture that pushes for innovation and being ahead of the curve, creating pressure for engaged teams to both take the time to question and explore and to deliver at the same time. Such pressure ensures that teams ultimately have to fall short on some aspects of their work, accruing new types of social and political costs for the teams. The teams must navigate increasing logistical and procedural trade-offs, such as spending more time and money on a project to accommodate additional work or changed features. In case of project course corrections, such costs can be political, in that the team needs to put extra effort into negotiating and convincing across layers of hierarchy. Project modifications or course corrections are sometimes not possible or feasible for design or development teams not involved in bigger decision making processes \cite{dindler2022engagements, Ma2024SenseOfMorality}. In these cases, teams may need to compromise on their own ethical standards, potentially resulting in additional moral stress \cite{reynolds2012moral}. 

Part of the challenge is that toolkits localise ethics within technical teams, towards the bottom of organisational structures. As a result, where the team in our study clearly increased their moral awareness, this was not happening evenly throughout the organization. Where ethics practices are anchored in individuals and small teams instead of organizations, ethics interventions fall short of impact and put the individuals into difficult and draining positions, when they must navigate the social costs these interventions impose \cite{raji2021you}. Widder and Nafus \cite{widder2023dislocated} discuss these relational factors as dislocated accountabilities, arguing that the organisational infrastructures necessary for technical development locate social appropriateness of what can be said, by whom, and with which results. 
In our study, the team was not being wilfully ignored in their concerns, yet they encountered the limitations of their roles in articulating their anxieties, as organizational structures are not built to attend to what Sarah in our study refers to as ''vague gut feelings'' when trying to estimate costs and utility benefits from technical solutions. 

These limitations are inherent to the nature of toolkits - toolkits try to make things accessible, with a low threshold for engagement \cite{wong2023seeing}. They therefore have to target small circles of impact in order to make them actionable and easily adoptable. Expecting toolkits to be localised anywhere beyond the individuals and small teams is opposite to their purpose. Toolkits can create the motivation and necessity for change, but they can not provide guidance for further necessary organisational changes that are required for outcomes such as moral stress to be absorbed into the relational systems of an organisation. Even if reflection work and ethical sensitivity can be developed from the bottom up, the management of the emotional cost needs to be acknowledged and accounted for from the top. Otherwise, when employees are left to deal with moral stress on their own without sufficient power to create meaningful change, teams run the risk of surrendering to the pressures of moral stress, potentially choosing ignorance and resignation as coping mechanisms \cite{corley2001development, imbulana2021interventions}. 

Ethics, in other words, is hard emotional labour and must be recognized as such, in order to provide meaningful options for stress mitigation. Ironically, engaging in it fully in this kind of labor requires recognition that this will add challenge and difficult to existing processes \cite{stark2009creative}. This is in contrast with common business practices seeking to avoid additional questions or uncertainties that rock the boat both within teams and outside of them \cite{stark2009creative, irani2019chasing}. It is clear that for ethics interventions to succeed, teams require emotional scaffolding in addition to logistical resources and the design of interventions must take notice of this requirement. Organizations must internalize that ethics considerations are tricky and projects will become more openly complicated when surfacing ethical tensions. Feminist scholars \cite{drage2024engineers} call for organisational cultures grounded in feminist notions of dynamic response-ability, a revaluation of care and maintenance practices and shared responsibilities. Yet even in best case scenarios organisational structures can be notoriously difficult to shift. Establishing explicit support structures and strategies for self-care, such as de-briefing sessions that explicitly give space to raise and share the felt experiences and affective impact of ethics work, and giving permission to admit and voice feelings of inadequacy and vulnerability, are some of ways to mitigate and address the emotional labour baked into technical work \cite{ballam2019emotionwork}. We hope that being able to identify moral stress as a core challenge can contribute to efforts aiming to design interventions that can provide the emotional scaffolding to technical teams grappling with ethical implications, and focus organisational attention to support the necessary within-team collaboration and care practices.  
 
\section{Conclusion}
In this paper, we discuss the affective experience of integrating ethics driven interventions in technical work practices. Even when toolkits are designed with logistical necessities in mind and technologists are emotionally invested and motivated to “do the right thing”, ethics interventions require emotional labour and produce moral stress, in the form of vulnerability, discomfort, and other difficult emotions. We argue that a successful and sustainable implementation of ethics initiatives requires accounting for inevitable moral stress \cite{reynolds2012moral}. Beyond the guidance on how to “do” ethics, teams require support in learning how to acknowledge and navigate the affective experiences they encounter, and receive reassurance that uncomfortable emotions are not necessarily a sign of wrong conduct, but an inevitable part of the experience \cite{imbulana2021interventions}.

Ethics interventions are often located within technical teams, without any expectations of shifts in organizational practice. While organizations might be willing to support their technical teams by providing extra time and resources for integrating new ethics-related tools and processes, few recognize the necessity for broader organizational change. We show that teams that embrace moral awareness and attempt to influence organizational decisions accordingly often encounter barriers in the form of managerial hierarchies and internal politics. Such barriers can lead to frustration, disengagement and stunted impact of any intervention that aims to develop moral awareness without political reach. Overlooking the inevitability of moral stress can explain why ethics materials meet little adoption or acceptance \cite{frauenberger2017action, madaio_co-designing_2020, metcalf2019owning}, especially where teams fully embrace the intentions and expectations of ethics initiatives. Working through ethical tensions therefore needs to be a full-scale organizational commitment, where moral stress is recognized as a legitimate part of ethical work through supportive structures. Going beyond typical organizational resources such as project management and communication structures, organizations must provide emotionally supportive scaffolding that acknowledges the extra costs of moral awareness and moral stress \cite{corley2001development, reynolds2015recognition}. 

In closing we want to emphasize that this work is not intended as a criticism for the particular organisation we worked with. Our feedback and findings about moral stress have been met with enthusiasm, where the organisation is willing to do the necessary work that is required in building in ethical reflection into technical practice. In fact, we believe that we were able to initially identify the manifestations of moral stress so clearly precisely because the team we collaborated with felt safe in openly voicing their doubts and concerns. We see further promising directions for this research in deepening the understanding of how ethics initiatives can address the socio-political dimensions attached to organizational practices and hope to motivate other researchers to follow a path that recognizes practitioners as emotional beings with affective needs when grappling with responsibilities, ethical tensions, and uncertain consequences.

\begin{acks}
We thank the teams that welcomed us to spend time with them and talk about the difficult topics of ethics in innovation practices. This work is part of the DCODE project, funded by the European Union’s Horizon 2020 research and innovation programme under the Marie Skłodowska-Curie grant agreement No 955990. 
This work has also been supported by the BUGGED project funded by the Research Council of Finland decision number: 348391 and the Nokia foundation's Jorma Ollila Grant. Views and opinions expressed are those of the authors only and do not necessarily reflect those of the funding organisations. 
\end{acks}

\bibliographystyle{ACM-Reference-Format}
\bibliography{moral-stress-ref}
\end{document}